\documentclass[conference]{IEEEtran}

\usepackage[utf8]{inputenc}
\usepackage{amsmath}
\usepackage{graphicx}
\usepackage{balance}

\usepackage{eso-pic}

\IEEEoverridecommandlockouts
\usepackage{cite}
\usepackage{amsmath,amssymb,amsfonts}
\usepackage{algorithmic}
\usepackage{graphicx}
\usepackage[nolist]{acronym}
\usepackage{textcomp}
\usepackage{multicol}
\usepackage{xcolor}
\usepackage{acronym}
\def\BibTeX{{\rm B\kern-.05em{\sc i\kern-.025em b}\kern-.08em
    T\kern-.1667em\lower.7ex\hbox{E}\kern-.125emX}}
\begin{document}

\title{Enhancing Cochlear Implant Signal Coding with Scaled Dot-Product Attention}

\author{\IEEEauthorblockN{Billel Essaid}
\IEEEauthorblockA{\textit{LSEA Lab., University of Medea}\\
Medea, 26000, Algeria \\
essaid.billel@univ-medea.dz}
\and
\IEEEauthorblockN{Hamza Kheddar}
\IEEEauthorblockA{\textit{LSEA Lab., University of Medea}\\
Medea, 26000, Algeria \\
kheddar.hamza@univ-medea.dz}
\and
\IEEEauthorblockN{Noureddine Batel}
\IEEEauthorblockA{\textit{LSEA Lab., University of Medea}\\
Medea, 26000, Algeria \\
batelnour@gmail.com}
}

\makeatletter

\def\ps@headings{%
\def\@oddhead{\parbox[t][\height][t]{\textwidth}{\flushleft

\noindent\makebox[\linewidth]
}
\vspace{0.5cm}
\hfil\hbox{}}%
\def\@oddfoot{\MYfooter}%
\def\@evenfoot{\MYfooter}}

\def\ps@IEEEtitlepagestyle{%
\def\@oddhead{\parbox[t][\height][t]{\textwidth}{
2024 International Conference on Telecommunications and Intelligent Systems (ICTIS)\\

}\hfil\hbox{}}%

\def\@oddfoot{ 979-8-3315-2739-6/24/\$31.00 \textcopyright 2024 IEEE \hfil 
\leftmark\mbox{}}%
\def\@evenfoot{\MYfooter}}

\maketitle

\maketitle

\begin{abstract}
 \Acp{CI} play a vital role in restoring hearing for individuals with severe to profound sensorineural hearing loss by directly stimulating the auditory nerve with electrical signals. While traditional coding strategies, such as the \ac{ACE}, have proven effective, they are constrained by their adaptability and precision. This paper investigates the use of \ac{DL} techniques to generate electrodograms for \acp{CI}, presenting our model as an advanced alternative. We compared the performance of our model with the \ac{ACE} strategy by evaluating the intelligibility of reconstructed audio signals using the \ac{STOI} metric. The results indicate that our model achieves a \ac{STOI} score of 0.6031, closely approximating the 0.6126 score of the \ac{ACE} strategy, and offers potential advantages in flexibility and adaptability. This study underscores the benefits of incorporating \ac{AI} into \ac{CI} technology, such as enhanced personalization and efficiency. 
\end{abstract}


\begin{table*}[]
    \centering
\begin{multicols}{3}
\footnotesize
\begin{acronym}[GCPFE]
\acro{CNN}{convolutional neural network}
\acro{AI}{artificial intelligent}
\acro{DL}{deep learning}
\acro{CI}{cochlear implant}
\acro{ML}{machine learning}
\acro{FCN}{fully convolutional neural networks}
\acro{STOI}{short time  objective intelligibility}
\acro{MMSE}{minimum mean square error}
\acro{DDAE}{deep denoising auto-encoder}
\acro{SNR}{signal to noise ratio}
\acro{EAS}{electric and acoustic stimulation}
\acro{CGRU}{ convolutional recurrent neural network with gated recurrent units}
\acro{CWT}{ continuous wavelet transform}
\acro{GRU}{gated recurrent units}
\acro{IGCIP}{image guided cochlear implant programming }
\acro{ESTOI}{extended short-time
objective intelligibility}
\acro{Bi-LSTM}{bidirectional long short term memory }
\acro{AAD}{auditory attention decoding}
\acro{SVM}{support vector machine}
\acro{CT}{computed tomography}
\acro{OCT}{optical coherence tomography}
\acro{MEE}{mean endpoint error}
\acro{TMHINT}{taiwan mandarin hearing in noise test}
\acro{Acc}{Accuracy}
\acro{Sen}{Sensitivity}
\acro{Spe}{Specificity}
\acro{Pre}{Precision}
\acro{Rec}{Recall}
\acro{F1}{F1 score}
\acro{NCM}{normalized covariance measure}
\acro{FFT}{fast Fourier transform}
\acro{ACE}{advanced combination encoder}
\acro{LGF}{loudness growth function}
\acro{MMS}{min-max similarity}
\acro{BF}{boundary F1}
\acro{ROC}{ receiver operating characteristic curve}
\acro{NCM}{normalized covariance measure}
\acro{DCS}{dice coefficient similarity}
\acro{ASD}{average surface distance}
\acro{AVD}{average volume difference}
\acro{JI}{jaccard index}
\acro{IoU}{intersection over union}
\acro{MAE}{mean absolute error}
\acro{HD}{hausdorff distance }
\acro{VAE}{variational autoencoders}
\acro{CAE}{convolutional autoencoder}
\acro{AE}{autoencoder}
\acro{SOTA}{state-of-the-art}
\acro{MFCC}{mel-frequency cepstral coefficient}
\acro{GFCC}{gammatone frequency cepstral coefficient }
\acro{AMR}{adaptive multi-rate}
\acro{LSTM}{long short term memory}
\acro{ASR}{automatic speech recognition}
\acro{DNN}{deep neural networks}
\acro{CS}{channel selection}
\acro{RNN}{recurrent neural network}
\acro{GAN}{generative adversarial network}
\acro{PESQ}{ perceptual estimation of speech quality}
\acro{KLT}{karhunen-loéve transform}
\acro{logMMSE}{log minimum mean squared error}
\acro{DAE}{deep autoencoder}
\acro{ICA}{intra cochlear anatomy}
\acro{ST}{scala tympani}
\acro{SV}{scala vestibul}
\acro{AR}{active region}
\acro{ASM}{active shape model}
\acro{NR}{noise reduction}
\acro{Res-CA}{residual channel attention}
\acro{GCPFE}{global context-aware pyramid feature extraction}
\acro{ACE-Loss}{active contour with elastic loss}
\acro{DS}{deep supervision}
\acro{UHR}{ultra-high-resolution}
\acro{ESCSO}{enhanced swarm-based crow search optimization}
\acro{cGAN}{conditional generative adversarial networks}
\acro{MAR}{metal artifacts reduction}
\acro{P2PE}{point to point error}
\acro{ASE}{average surface error}
\acro{MARGAN}{metal artifact reduction based generative adversarial networks}
\acro{ESCSO}{ enhanced swarm based crow search optimization}
\acro{EEG}{electroencephalography}
\acro{ERP}{event-related potential}
\acro{ANN}{artificial neural network}
\acro{RBNN}{radial basis functions neural networks}
\acro{KNN}{K-nearest neighbours}
\acro{RF}{random forests}
\acro{DRNN}{deep recurrent neural networks}
\acro{MLP}{multilayer perceptrons}
\acro{NMF}{non-negative matrix factorization}
\acro{DCAE}{deep convolutional auto-encoders}
\acro{ILD}{interaural level difference}
\acro{RTF}{relative transfer function}
\acro{SDR}{source-to-distortion ratio}
\acro{SAR}{source-to-artifact ratio}
\acro{SIR}{source-to-interference ratio}
\acro{dB}{decibel}
\acro{MIMO}{multiple-input multiple-output}
\acro{EVD}{eigenvector decomposition}
\acro{ReLU}{rectified linear unit}
\acro{RNN}{recurrent neural network}
\acro{MICE}{multiple imputation by chained equations}
\acro{ECochG}{intracochlear electrocochleography}
\acro{CM}{cochlear microphonic}
\acro{FFR}{frequency following responses}
\acro{$f_0$}{fundamental frequency}
\acro{TFSC}{temporal fine structure cues}
\acro{RMSE}{root mean square error}
\acro{BM}{basilar-membrane}
\acro{ROI}{regions of interest}
\acro{DRL}{deep reinforcement learning}
\acro{RL}{reinforcement learning}
\acro{ABR}{auditory brainstem responses}
\acro{DBS}{deep brain stimulation}
\acro{FOX}{fitting to outcomes expert}
\acro{DSP}{digital signal processor}
\acro{MHINT}{mandarin of hearing in noise 
test}
\acro{BCP}{bern cocktail party}
\acro{EST}{effective stimulation threshold}
\acro{ECAP}{electrically evoked compound action potential}
\acro{MRI}{magnetic resonance 
imaging}
\acro{BSS}{blind source separation}
\acro{AEP}{auditory evoked potential}
\acro{CCA}{canonical correlation analysis}
\acro{ASE}{average surface error}
\acro{STOI}{short-time objective intelligibility}
\acro{TL}{transfer learning}
\acro{CIS}{continuous interleaved sampling}
\acro{DTL}{deep transfer learning}
\acro{DT}{decision tree}
\acro{FL}{federated learning}
\acro{CTC}{connectionist temporal classification}
\acro{BERT}{bidirectional encoder representations from transformer}
\acro{CI}{cochlear implant}
\acro{TCN}{temporal convolutional network}
\acro{MSE}{mean squared error}
\acro{BCE}{binary cross entropy}

\end{acronym}

\end{multicols}
\end{table*}

\section{Introduction}

\acp{CI} are advanced medical devices designed to restore hearing for individuals with severe to profound sensorineural hearing loss. These implants bypass damaged or nonfunctional parts of the inner ear by directly stimulating the auditory nerve with electrical signals, enabling users to perceive sound. Traditional coding strategies, such as the \ac{ACE}, have been widely adopted to convert acoustic signals into electrical stimulation patterns for \acp{CI}. While these methods have proven effective, they are limited in their ability to adapt to diverse speech environments and individual variations in auditory processing.

In recent years, the rise of \ac{AI} and \ac{DL} has opened new avenues for improving biomedical application \cite{habchi2023ai}, speech recognition, specifically  \ac{CI} coding strategies \cite{essaid2024artificial}. By leveraging \ac{AI}, it is possible to develop models that not only mimic traditional coding methods but also enhance the adaptability, efficiency, and accuracy of signal processing. \ac{AI}-driven models, have the potential to capture more complex patterns in speech and deliver more personalized auditory experiences to \ac{CI} users.

This paper explores the application of \ac{DL} in generating stimulation signals for \acp{CI}, comparing the performance of an \ac{AI}-based model to the traditional \ac{ACE} strategy. By evaluating both approaches using objective metrics such as the \ac{STOI} score, we demonstrate the potential of \ac{AI} to transform the field of \acp{CI}, improving both the precision and adaptability of auditory signal processing. 

In this work, we integrate \ac{TCN} and the scaled dot-product attention mechanism \cite{kheddar2024transformers} to enhance the \ac{CI} coding strategy. This approach leverages \ac{TCN}’s ability to model long-range dependencies in sequential data, while attention focuses on key parts of the input signal. Together, these techniques improve the accuracy and adaptability of electrodogram generation compared to traditional methods like \ac{ACE}.

The structure of this paper is organized as follows: Section II provides background information on \acp{CI} and their traditional coding strategies. Section III explores how \ac{AI} can enhance \ac{CI} coding, focusing on modern approaches. In Section IV, we present the proposed method, detailing the integration of \ac{TCN} and scaled dot-product attention. Finally, Section V presents the experimental results, including a comparison of our model with the traditional \ac{ACE} strategy, demonstrating its performance and effectiveness.

\section{CIs Background}

\subsection{CIs principle}

\begin{figure}[h]
    \centering
    \includegraphics[scale=0.33]{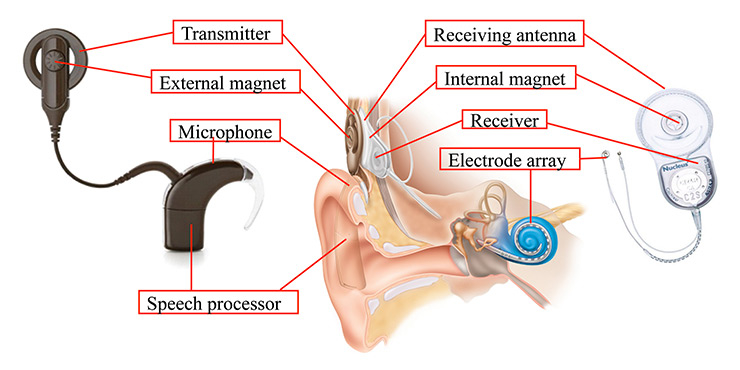}
    \caption{Diagram of the components of a \ac{CI}. \cite{brand2014cochlear}.}
    \label{fig:cohlear}
\end{figure}

\acp{CI} are a common treatment for individuals with severe to profound sensorineural hearing loss. These devices directly stimulate the auditory nerve with electrical signals, bypassing the inner ear's damaged or absent hair cells. This electrical stimulation enables users to perceive sound and restore a degree of hearing. A \ac{CI} consists of several key components that work together to enable sound perception as illustrated in Fig.\ref{fig:cohlear} \cite{brand2014cochlear}. 

The external part includes a microphone, which picks up sound from the environment, and a speech processor that converts these sounds into digital signals. These signals are then sent to a transmitter coil \cite{essaid2016evaluation}, which sits on the scalp and transmits the information to the internal components through radio frequency waves. The internal part of the \ac{CI} consists of a receiver/stimulator implanted under the skin, which decodes the signals and sends them to an array of electrodes placed in the cochlea. These electrodes directly stimulate the auditory nerve, bypassing the damaged hair cells, and allowing the brain to interpret the signals as sound \cite{essaid2018new}.

\subsection{Stimulation strategies in CIs}

\begin{figure}[h]
    \centering
    \includegraphics[scale=0.47]{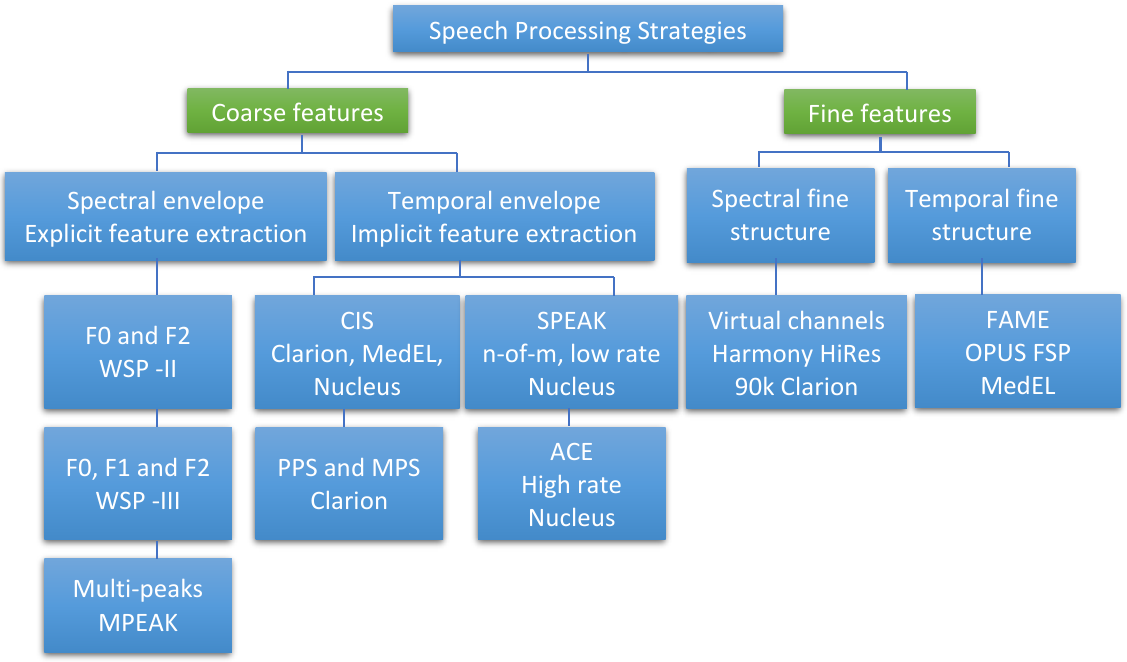}
    \caption{Categories of signal processing strategies in CIs \cite{zeng2008cochlear}.}
    \label{fig:strategies}
\end{figure}

Stimulation strategies are pivotal in maximizing the efficacy of \acp{CI}, with various techniques tailored to enhance auditory outcomes for users as detailed in Fig. \ref{fig:strategies}.

\begin{figure}[h]
    \centering
    \includegraphics[scale=0.48]{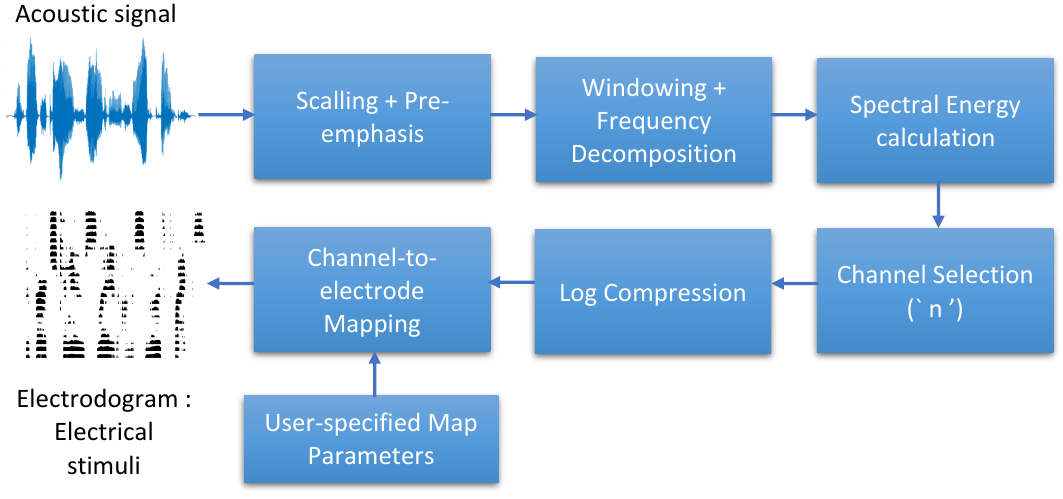}
    \caption{Basic block diagram of ACE processing strategy \cite{hansen2019cci}}.
    \label{fig:ACE}
\end{figure}

Among these strategies, the \ac{ACE} stands out as a sophisticated algorithmic approach that aims to provide clear and natural auditory information to implant recipients. By strategically distributing frequency information across the electrode array in the cochlea as presented in Fig. \ref{fig:ACE}, \ac{ACE} focuses on optimizing speech perception by improving the clarity and discrimination of speech sounds. 

Complementing \ac{ACE}, the \ac{CIS} strategy prioritizes temporal cues in speech processing by delivering electrical stimulation through individual electrodes sequentially. Conversely, the spectral peak (SPEAK) strategy accentuates spectral peaks in speech signals to refine frequency resolution and bolster speech understanding. 
Research indicates that the selection and optimization of these strategies are pivotal in achieving optimal outcomes for \ac{CI} users, highlighting the importance of individualized approaches to maximize the benefits of these remarkable devices.

\section{Enhancing CI coding strategies through AI}

This section discusses advancements in \ac{CI} coding strategies through \ac{AI}, covering the integration of \ac{AI} into stimulation strategies and the role of \ac{TCN} and scaled dot-product attention in speech signal processing.

\subsection{AI and CI stimulation strategies}

\ac{AI} has emerged as a cutting-edge tool in optimizing coding strategies for \acp{CI}, revolutionizing the field of auditory prosthetics. By harnessing the power of \ac{ML} algorithms, \ac{AI} can analyze vast amounts of data to tailor stimulation strategies to individual \ac{CI} users. These \ac{AI}-driven coding strategies aim to enhance speech perception, refine sound processing, and personalize auditory experiences based on the unique needs of each recipient. Through advanced pattern recognition and adaptive learning capabilities, \ac{AI} algorithms can dynamically adjust stimulation parameters to optimize speech understanding and maximize communication outcomes. The integration of \ac{AI} in \ac{CI} coding not only streamlines the customization process but also holds the potential to significantly improve the effectiveness and efficiency of \ac{CI} programming. As research in this domain continues to evolve, the synergy between \ac{AI} and \acp{CI} promises to open new frontiers in enhancing auditory rehabilitation and transforming the lives of individuals with hearing loss \cite{essaid2024artificial}.

Integrating artificial intelligence, particularly deep learning, in speech coding strategies for \acp{CI} represents a significant advancement in hearing restoration technology. These sophisticated algorithms are being employed to enhance signal processing, improve speech recognition, and optimize the overall auditory experience for \ac{CI} users. \ac{DL} models, particularly \acp{CNN} and \acp{RNN}, are being utilized to analyze complex acoustic features and extract relevant information from speech signals more effectively than traditional methods. This \ac{AI}-driven approach allows for more precise mapping of acoustic inputs to electrode stimulation patterns, potentially leading to improved speech understanding and sound quality perception. Additionally, \ac{ML} techniques are being explored to personalize \ac{CI} settings based on individual user preferences and hearing profiles, potentially reducing the need for manual adjustments and enhancing overall user satisfaction.

Gajecki et al. \cite{gajecki2022end,gajecki2023deep}, propose a \ac{DL} speech denoising sound coding strategy that estimates the \ac{CI} electric performing end-to-end \ac{CI} processing. The results show that the proposed method is capable of replacing a \ac{CI} sound coding strategy while preserving its general use for every listener and performing speech enhancement in noisy environments without sacrificing algorithmic latency. The model was optimized using a \ac{CI}-specific loss function and evaluated using objective measures, showing higher scores than baseline algorithms. This study introduces a novel \ac{DL} approach,  which has the potential to significantly improve speech intelligibility for \ac{CI} users in challenging acoustic environments.

Furthermore, Huang et al. \cite{huang2023electrodenet}, proposes a \ac{DL}-based sound coding strategy for \acp{CI}s, which emulates the \ac{ACE} strategy by replacing the conventional processing stages with a \ac{DNN}. The extended ElectrodeNet-CS strategy further incorporates channel selection. The network models of \ac{DNN}, \ac{CNN}, and \ac{LSTM} were trained using the Fast Fourier Transformed bins and channels. Objective speech understanding using \ac{STOI} and normalized sentence recognition tests for vocoded Mandarin speech were conducted with normal-hearing listeners. DNN, CNN, and LSTM-based ElectrodeNets exhibited strong correlations to ACE in objective and subjective scores. The methods and findings demonstrated the feasibility and potential of using \ac{DL} in \acp{CI}.

\subsection{\Ac{TCN} and scaled dot-product attention mechanisms in speech processing}

The integration of \ac{TCN} and scaled dot-product attention mechanisms in speech processing has significantly enhanced the capabilities of various applications, particularly in speech enhancement and emotion recognition.

\acp{TCN} are designed to capture long-range dependencies in sequential data, making them particularly suitable for tasks involving speech signals. Unlike \acp{RNN}, which process data sequentially, \acp{TCN} leverage convolutional layers to facilitate parallel processing, thus improving computational efficiency. \acp{TCN} utilize causal convolutions, ensuring that the model only has access to past information, which is crucial for real-time applications like speech enhancement. Studies have shown that \acp{TCN} can outperform \acp{RNN} in modelling long-term dependencies, especially in noisy environments, where accurately distinguishing speech from background noise is essential \cite{nicolson2020masked,essaid2025deep}.

Given an input sequence \( x = (x_1, x_2, \dots, x_T) \), the output of a \ac{TCN} is produced by multiple layers of 1D convolutions. For each layer, the output at time \( t \), denoted as \( y_t \), is computed using the convolutional filter over past input values as defined in equation \eqref{eq1}:

\begin{equation}
y_t = f(W * x_{t-k:t}, b), 
\label{eq1}
\end{equation}

\noindent where:
\begin{itemize}
    \item \( W \) represents the learned convolutional filters.
    \item \( x_{t-k:t} \) is the segment of the input sequence being convolved.
    \item \( * \) denotes the convolution operation.
    \item \( f \) is an activation function, such as ReLU or Tanh.
    \item \( b \) is the bias term.
\end{itemize}

In dilated convolutions, the receptive field grows with depth, meaning the network can learn to model dependencies over a longer time horizon.

The scaled dot-product attention mechanism is a fundamental component of the Transformer architecture \cite{kheddar2024transformers,djeffal2023automatic}, which has been successfully adapted for various speech processing tasks. This attention mechanism computes the relevance of different parts of the input sequence by calculating the dot product of query and key vectors, followed by a softmax operation to obtain attention weights. The scaling factor, $\sqrt{d_k}$, where $d_k$ is the dimensionality of the key vectors, is used to stabilize the gradients during training, preventing the softmax function from saturating \cite{vaswani2017attention}. This mechanism allows the model to focus on specific segments of the input, enhancing its ability to capture contextual relationships within speech data.

The mathematical formulation of the scaled dot-product attention is given equation \eqref{eq2} \cite{kheddar2024automatic,djeffal2024transformer}:

\begin{equation}
\text{Attention}(Q, K, V) = \text{softmax}\left(\frac{QK^T}{\sqrt{d_k}}\right)V
\label{eq2}
\end{equation}

where:
\begin{itemize}
    \item \( Q \in \mathbb{R}^{n \times d_k} \) is the matrix of queries,
    \item \( K \in \mathbb{R}^{m \times d_k} \) is the matrix of keys,
    \item \( V \in \mathbb{R}^{m \times d_v} \) is the matrix of values,
    \item \( d_k \) is the dimensionality of the key vectors,
    \item \( \frac{1}{\sqrt{d_k}} \) is the scaling factor that normalizes the dot product.
\end{itemize}

In this work, we leverage the strengths of both \acp{TCN} and the scaled dot-product attention mechanism to improve speech processing tasks in \acp{CI}. \acp{TCN} excel at capturing local temporal dependencies through causal and dilated convolutions, while attention mechanisms dynamically capture global relationships by focusing on the most relevant parts of the input sequence. By combining these approaches, our model benefits from the parallel processing and noise robustness of \acp{TCN}, alongside attention's flexibility and contextual understanding. This hybrid architecture enhances the model’s ability to efficiently process complex speech patterns and long-range dependencies, improving speech-related tasks' performance.

\section{Proposed method}

This section details our \ac{DL} model, describes the audio materials used for training, and explains the \ac{STOI} metric employed to evaluate the model's performance.

\subsection{\ac{DL} model}

\begin{figure}[h]
    \centering
    \includegraphics[scale=0.54]{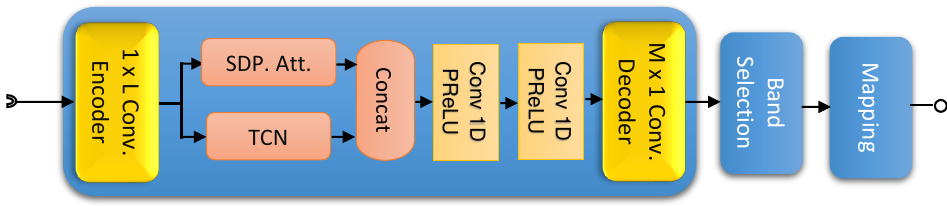}
    \caption{Structure of the proposed model. }
    \label{fig:model}
\end{figure}

The proposed model presented in Fig. \ref{fig:model} embodies an innovative fusion of cutting-edge technologies, notably the \ac{TCN} operating in tandem with scaled dot product attention, promising a paradigm shift in \ac{CI} coding strategies. The Encoder module serves as the entry point for the input data, extracting crucial features that are essential for subsequent stages of processing, and at the heart of this model lies the \ac{TCN}, a dynamic neural network component finely tuned to unravel intricate temporal dependencies within auditory data. This strategic pairing with scaled dot product attention introduces a novel dimension of information processing, enabling the model to discern nuanced relationships across data points, thereby enhancing feature integration and representation.

The synergy between \ac{TCN} and scaled dot product attention within the model heralds a new era in auditory signal processing. The \ac{TCN}'s proficiency in capturing temporal nuances harmonizes seamlessly with the attention mechanism's ability to spotlight critical data correlations, fostering a comprehensive understanding of auditory signals. This collaborative approach not only refines the model's feature extraction capabilities but also enriches the auditory experience by amplifying speech intelligibility and enhancing sound clarity.

\subsection{Audio material}

For the development and evaluation of the model, the TIMIT database was employed as the primary data source. Each .wav audio file was converted into an electrodogram using the \ac{ACE} strategy implemented in Matlab, simulating the electrical stimulation patterns used in \acp{CI} to represent sound. The dataset consisted of 100 audio files for both training and validation, with corresponding electrodograms as the target outputs. The files used in validation were unseen during training to ensure optimal performance. Additionally, 20 files were reserved for testing to guarantee a comprehensive evaluation. The data was meticulously pre-processed to maintain the fidelity of the electrodogram representations and ensure compatibility with the model architecture. A combined loss function, incorporating both \ac{MSE} and \ac{BCE}, was used to optimize the model, serving distinct roles: \ac{MSE} handles regression by minimizing the difference between predicted and actual electrodogram signals to ensure accurate signal reconstruction, while \ac{BCE} supports classification by guiding the selection of relevant frequency bands. This combined approach enables the model to enhance both signal fidelity and functional auditory clarity. This approach allowed for a more nuanced evaluation, capturing both regression and classification aspects of the prediction task. The setup facilitated robust model optimization across diverse speech patterns and acoustic features \cite{kheddar2023deep}.

\subsection{Objective metric}

A vocoder was employed to convert the predicted electrodograms back into audio signals, enabling auditory comparison between the original and reconstructed signals. The quality of these reconstructed signals was assessed using the \ac{STOI} metric. 

\ac{STOI} metric provides an objective evaluation of speech intelligibility by comparing the temporal envelopes of clean and degraded speech signals. The process begins by dividing both the clean signal $s[n]$ and the degraded signal $\hat{s}[n]$ into overlapping short-time frames, followed by applying the Short-Time Fourier Transform (STFT) to extract their time-frequency representations. These are then filtered into perceptually relevant third-octave bands, yielding temporal envelopes $s(l, k)$ and $\hat{s}(l, k)$ for each frame $l$ and frequency band $k$. The core of STOI is the computation of the correlation coefficient $\rho(l, k)$ between the clean and degraded envelopes for each band and frame.
The final \ac{STOI} score is obtained then by averaging the correlation across all frames $L$ and frequency bands $K$ \cite{taal2010short}:

\begin{equation}
    \text{STOI} = \frac{1}{LK} \sum_{l=1}^{L} \sum_{k=1}^{K} \rho(l, k)
    \label{eq3}
\end{equation}

This score, ranging from 0 to 1, indicates the intelligibility of the degraded speech, with 1 representing perfect intelligibility and 0 indicating unintelligible speech.

\section{Experimental results}

The training process is carefully managed using early stopping and adaptive learning rate strategies to enhance performance and prevent overfitting. Starting with an initial learning rate of $1 \times 10^{-3}$, the training runs for up to 300 epochs, but early stopping with a patience of 5 epochs halts training if validation loss does not improve, avoiding unnecessary computations. Additionally, the learning rate is reduced by 20\% if the validation loss stagnates for 3 epochs, ensuring efficient learning throughout. The best model is saved based on the lowest validation loss, guaranteeing optimal performance while minimizing overfitting.

\begin{figure}[h]
    \centering
\includegraphics[scale=0.55]{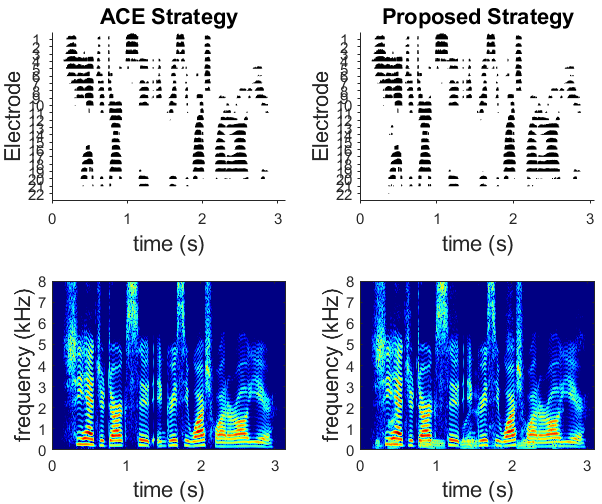}
    \caption{Electrodogram generation and vocoded signals comparaison.}
    \label{fig:electrodogram}
\end{figure}

Fig. \ref{fig:LCurve} demonstrates the training and validation curves obtained after the training process.:

\begin{figure}[h!]
    \centering
\includegraphics[height=6cm, width=9cm]{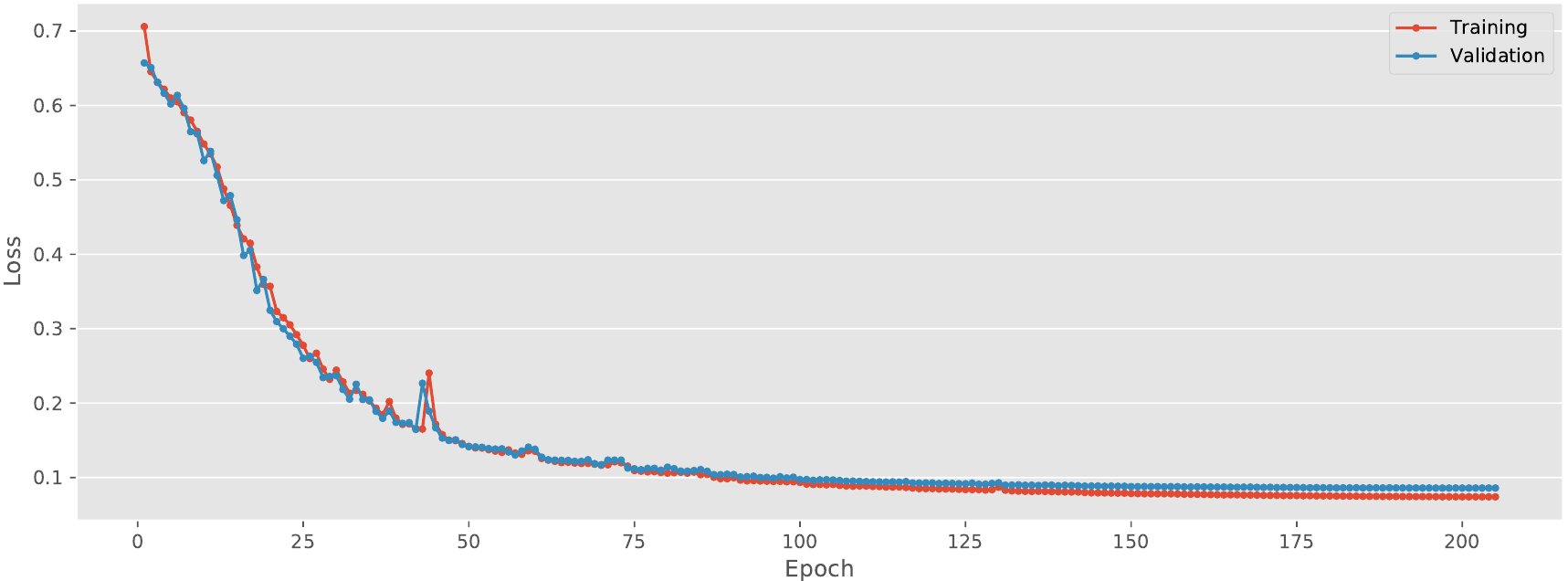}
    \caption{Training and validation performance}.
    \label{fig:LCurve}
\end{figure}
After saving the best-performing model, it was subsequently used for testing. An audio signal from the test set was input into the model, and the corresponding electrodogram was predicted. To evaluate the model's effectiveness, the predicted electrodogram was compared against the electrodogram generated by the \ac{ACE} coding strategy, which serves as the ground truth. Fig. \ref{fig:electrodogram} illustrates this comparison, presenting both the electrodogram produced by our model and the one generated by the \ac{ACE} strategy. This side-by-side comparison allows for a visual assessment of the model’s ability to replicate the complex electrical stimulation patterns essential for \ac{CI} function, highlighting its potential in accurately mimicking the auditory processing performed by the \ac{ACE} strategy. Such comparisons are crucial for demonstrating the model's precision and effectiveness in real-world applications.

Fig. \ref{fig:electrodogram} presents also the spectrograms of the audio signals generated by both the \ac{ACE} strategy and our model, obtained using a sine wave vocoder. The spectrograms offer a detailed visual representation of the frequency content over time, enabling a comprehensive comparison between the two signals. To quantitatively assess the intelligibility of the reconstructed audio, the \ac{STOI} metric was employed. The \ac{STOI} metric provides an objective measure of speech intelligibility, allowing us to compare the performance of our model against the \ac{ACE} strategy. This combination of visual and quantitative analysis helps demonstrate the model's capability to generate high-quality audio signals that closely resemble those produced by the \ac{ACE} strategy, particularly in terms of preserving critical speech characteristics and intelligibility.

After applying the test dataset to our model, the predicted electrodograms were vocoded back into audio signals to assess their intelligibility. We calculated the average \ac{STOI} score for both the audio signals generated by our model and those produced by the traditional \ac{ACE} strategy. The results showed an average \ac{STOI} score of 0.6031 for our model and 0.6126 for the \ac{ACE} strategy.

While the \ac{ACE} strategy slightly outperforms our model in terms of intelligibility, the close proximity of the two scores indicates that our model is highly effective at replicating the auditory signals generated by the \ac{ACE} coding strategy. The minor difference in \ac{STOI} suggests that our model preserves the essential speech features and intelligibility to a comparable degree, demonstrating its potential as a viable alternative for generating electrodograms in \acp{CI}. This result is promising, as it highlights the ability of our model to approach the performance of the well-established \ac{ACE} strategy while offering the flexibility of a \ac{DL}-based approach.

In this study, our model achieved a \ac{STOI} slightly below the \ac{ACE} baseline using the TIMIT dataset. When compared to existing studies that utilized the TMHINT dataset, STOI values for ACE ranged from 0.678 \cite{huang2021combination} to 0.718 \cite{huang2023electrodenet}, with models like ElectrodeNet and its enhanced version achieving 0.707 and 0.717, respectively \cite{huang2023electrodenet}. The envelope enhancement (EE) strategy, particularly when combined with biologically-inspired hearing aid algorithm (BioAid), showed the highest STOI of 0.679 in TMHINT tests \cite{huang2021combination}. These differences may stem from the use of different datasets—TIMIT versus TMHINT—and variations in vocoders across studies, which can influence intelligibility outcomes. Despite these variations, our model demonstrates competitive performance within the context of \ac{CI} coding strategies.

\section{Conclusion}


In conclusion, this study demonstrates the potential of \ac{DL} to enhance \ac{CI} coding by introducing a novel model that integrates \ac{TCN} with scaled dot-product attention. Compared to the traditional ACE strategy, the proposed model offers competitive performance in speech intelligibility, measured by \ac{STOI} scores, while also providing greater flexibility and adaptability to diverse speech environments. The model’s ability to process complex auditory signals more accurately illustrates its value as a robust alternative for \ac{CI} users, promising a more personalized and effective auditory experience.

Future work could explore improving AI models with larger and more diverse datasets, optimizing architectures, and enabling real-time processing. Additionally, transfer learning \cite{kheddar2023deep,habchi2024deep,lachenani2024improving,djeffal2024transfer} could be leveraged to adapt models from similar tasks, enhancing flexibility and effectiveness in \ac{CI} technology. Integrating federated learning \cite{himeur2023federated} between right and left \acp{CI}  with real-world user feedback may lead to more adaptive, patient-specific solutions, ultimately improving outcomes for individuals in various acoustic environments.

\balance
\bibliographystyle{elsarticle-num}
\bibliography{references}

\end{document}